# A New Model-Free Method Combined with Neural Networks for MIMO Systems

Feilong Zhang

***Abstract*—In this brief, a model-free adaptive predictive control (MFAPC) is proposed. It outperforms the current model-free adaptive control (MFAC) in solving the time delay problem in multiple-input multiple-output (MIMO) systems and in relaxing the current rigorous assumptions for a wider application range. The most attractive merit of the proposed controller is that the controller design, performance analysis, and applications are easy for engineers to realize. Furthermore, the problem of choosing the matrix $\lambda$ is resolved by analyzing the function of the closed-loop poles rather than the previous contraction mapping method. Additionally, considering the nonlinear modeling capability and adaptability of neural networks (NNs), we combine these two classes of algorithms. The feasibility and several interesting results of the proposed method are demonstrated in simulations.**

## I. INTRODUCTION

The theses about MFAC have been extensively published during the past decade. The MFAC has a simplified discrete-time form for being suitably performed by computers, and it has been tested in process control, urban traffic, power systems, and so on [1]-[20]. The notion of "model-free" control is that the only requirement of controller design is not to build the system model but the I/O data of systems. This ascribes to the that the controller is determined by one process model termed an equivalent-dynamic-linear-model (EDLM) whose coefficients derive from the online estimated pseudo-gradient (PG) vector or pseudo-Jacobian matrix (PJM). However, many conclusions contradict [21]-[24].

Until now, the majority of literature about MFAC has focused on single-input and single-output systems (SISO). However, fewer are related to multivariable systems, and the current system descriptions are assumed with considerably rigorous restrictions. This mainly ascribes to the cross-coupling problem and the fact that the matrix multiplication normally does not satisfy commutativity. Furthermore, the different time delays among I/O variables always cause difficulties in controller design. To address this issue, [25] imposes the non-diagonal interactor matrix in general minimum variance control [26]. However, the interactor matrix is necessary to be ascertained by the preliminary experiment. In addition, as a typical representative in this kind of adaptive controller design, [27] has systematically introduced several pseudo exchange matrices (PEM) and interactor matrices, which generally results in more difficulties in controller designs and applications for engineers. On the other hand, the decoupling controller has received the most attention in the field of multivariable systems and its decoupling effect is hardly exempted from the restriction of modeling results in most occasions [28], [29]. In consequence, the decompiling controller may have more difficulties in applications compared to the directly designed controller in [27] and the proposed method in this brief. Moreover, the proposed method is more suitable for situations where the number of system inputs is inherently more than the number of system outputs, which naturally or intentionally helps to acquire a better performance.

The essence of EDLM is to express the nonlinear system model by the linearization around each operating point. Based on EDLM, the current MFAC is obtained by minimizing the quadratic index function. Therefore, it is intrinsically based on the linear model. More precisely, the adaptability to uncertainty or nonlinearity is made possible by introducing the online estimate algorithm in light of the certainty equivalence adaptive control in [25]-[27], which claims that someone can take one online estimate algorithm and one existing unfalsified controller to create a new scheme. However, to the best of the author's knowledge and generally speaking, every current estimate algorithm may have its own merits and suffers from one or more of its drawbacks. The selection principle can exert its advantages and avoid its weaknesses in view of the actual situations. Up to now, tremendous works have been proposed on the combination of neural networks (NNs) and the conventional adaptive controller in nonlinear systems for the claimed merit that any continuous function can be approximated within any desired accuracy for its powerful nonlinear modeling capability and adaptability [16]-[20], [30]-[33]. According to the chain rule of partial derivatives, we identify the elements in PJM by BPNN and RBFNN instead of the preferred least squares algorithm or projection method in this paper. Furthermore, their corresponding curves are exhibited for the first time and in an extraordinary style.

The main contributions of this brief are summarized as follows:

1) Current multivariable EDLM has limitations in objectively reflecting the actual system. One reason is that its leading coefficient matrix of the control input vector is assumed to be a diagonally dominant matrix, and another is that its pseudo-order is confined to $1 \leq L \leq n_u$. To make up for these deficiencies, the corrected EDLM in this paper is no longer restricted to the diagonally dominant square matrix and is relaxed with the different number of input and output variables. Additionally,

Manuscript received Dec 3, 2020. This work was supported in part by the xxxxxxx.

Feilong Zhang is with the State Key Laboratory of Robotics, Shenyang Institute of Automation, Chinese Academy of Sciences, Shenyang 110016, China (e-mail: zhangfeiong@sia.cn).

the permissible range of pseudo-order extends to $1 \le L$. The poof is formulated in Appendix I.

2) We predict the outputs of EDLM over several steps to put forward the MIMO-MFAPC which aims to minimize the quadratic performance function (8). One of the most significant advantages of the proposed method is that it's easy to design and implement the controller for engineers. One reason is that the interactor matrix and pseudo exchange matrix are not required for the controller design and the time delay problem is generally solved by choosing a proper predictive step $N$ which exceeds the maximum time delay. Contrariwise, any time delay may give rise to the failure of the current MFAC, even if all the coefficient matrices in the controller are substituted by the corresponding actual values of linear multivariable systems. (This is illustrated in [21]) On the other hand, this brief also strengths that the objective of MFAPC or current MFAC designed via partial-form EDLM is not based on unstable systems.

3) We finish the stability analysis of the proposed method and simultaneously present how to quantitatively choose $\boldsymbol{\lambda}$ or analyze the chosen $\boldsymbol{\lambda}$ by performance analysis. Whereas the conclusion distinguishes from the previous conclusion in [1]-[20] that $\lambda$ is large enough to ensure the convergence of tracking error.

4) According to the definition of derivative and principle of neural networks, we attempt to estimate the parameters of EDLM by NNs and obtain a remarkable result. On the other hand, the adjustment of $\boldsymbol{\lambda}$ through BPNN manifests that some negative elements in the weight matrix $\boldsymbol{\lambda}$ are compatible.

The rest of the paper is organized as follows. In Section II, the EDLM is proposed for a kind of multivariable nonlinear system. In Section III, the MFAPC design and its performance analysis are given. Then, the NNs are introduced as the estimate algorithm. In Section IV, a series of simulations are performed to show the feasibility and prospect of this kind of composite method. The conclusion is given in Section V. At last, the proof of corrected EDLM is attached in Appendix I.

## II. Description of Systems and Prediction Model

### A. Systems Description

Let the discrete-time MIMO nonlinear system be described by

$$\boldsymbol{y}(k+1) = \boldsymbol{f}(\boldsymbol{y}(k),\cdots,\boldsymbol{y}(k-n_y),\boldsymbol{u}(k),\cdots,\boldsymbol{u}(k-n_u)) \tag{1}$$

or the form of (2)

$$\begin{aligned} \boldsymbol{y}(k+1) &= \boldsymbol{f}(\boldsymbol{\varphi}(k)) \\ \boldsymbol{\varphi}(k) &= [\boldsymbol{u}(k),\cdots,\boldsymbol{u}(k-n_u)] \end{aligned} \tag{2}$$

where $\boldsymbol{f}(\cdots) = \left[f_1(\cdots),\cdots,f_m(\cdots)\right]^T$ is a continuous and smooth nonlinear vector-valued function with respect to all variables. $n_y+1$, $n_u+1 \in \mathrm{Z}$ are the orders of output vector $\boldsymbol{y}(k)$ and input vector $\boldsymbol{u}(k)$ of the system at operating point $k$, respectively. The dimensions of $\boldsymbol{y}(k)$ and $\boldsymbol{u}(k)$ are $M_y$ and $M_u$, respectively.

*Assumption 1:* The partial derivatives of $\boldsymbol{f}(\cdots)$ in (1) or (2) with respect to all vectors are continuous.

*Theorem 1*: Given system (1) or (2) such that the above assumption, if $\Delta \boldsymbol{U}(k) \ne \boldsymbol{0}$, $1 \le L$, there must exist a $\boldsymbol{\phi}_L^T(k)$ termed pseudo-Jacobian matrix and (1) or (2) can be reformulated as:

$$\Delta \boldsymbol{y}(k+1) = \boldsymbol{\phi}_L^T(k)\Delta \boldsymbol{U}(k) \tag{3}$$

where

$\boldsymbol{\phi}_L^T(k) = [\boldsymbol{\Phi}_1(k),\cdots,\boldsymbol{\Phi}_L(k)]$;

$\boldsymbol{\Phi}_i(k) \in \boldsymbol{R}^{m\times n}$, $m\times n = M_y \times M_u$, ($i$= 1,⋯, $L$);

$\Delta \boldsymbol{U}(k) = [\Delta \boldsymbol{u}^T(k),\cdots,\Delta \boldsymbol{u}^T(k-L+1)]^T$ is a vector that contains the increment of control input vector within $[k-L_u+1,k]$. The integer $1 \le L$ is the pseudo order.

Define $\boldsymbol{\phi}_L(z^{-1}) = \boldsymbol{\Phi}_1(z^{-1}) + \cdots + \boldsymbol{\Phi}_L(k)z^{-L+1}$.

### B. Prediction Model

It is easy to rewrite (3) as

$$\boldsymbol{y}(k+1) = \boldsymbol{y}(k) + \boldsymbol{\phi}_L^T(k)\Delta \boldsymbol{U}(k) \tag{4}$$

Herein, define

$$\boldsymbol{A} = \begin{bmatrix} \boldsymbol{0} & & & \\ \boldsymbol{I} & \boldsymbol{0} & & \\ & \ddots & \ddots & \\ & & \boldsymbol{I} & \boldsymbol{0} \end{bmatrix}_{(L\bullet M_u)\times(L\bullet M_u)} \qquad \boldsymbol{B}^T = \begin{bmatrix} \boldsymbol{I} & \boldsymbol{0} & \cdots & \boldsymbol{0} \end{bmatrix}_{M_u \times (L\bullet M_u)}$$

Following [25], a finite $N$-step forward prediction on the basis of (4) is given as

$$\begin{aligned} \boldsymbol{y}(k+1) &= \boldsymbol{y}(k) + \boldsymbol{\phi}_L^T(k)\Delta \boldsymbol{U}(k) \\ &= \boldsymbol{\phi}_L^T(k)\boldsymbol{A}\Delta \boldsymbol{U}(k-1) + \boldsymbol{\phi}_L^T(k)\boldsymbol{B}\Delta \boldsymbol{u}(k) \\ \boldsymbol{y}(k+2) &= \boldsymbol{y}(k+1) + \boldsymbol{\phi}_L^T(k+1)\Delta \boldsymbol{U}(k+1) \\ &\vdots \\ \boldsymbol{y}(k+N) &= \boldsymbol{y}(k+N-1) + \boldsymbol{\phi}_L^T(k+N-1)\Delta \boldsymbol{U}(k+N-1) \\ &= \boldsymbol{y}(k) + \sum_{i=0}^{N-1}\boldsymbol{\phi}_L^T(k+i)\boldsymbol{A}^{i+1}\Delta \boldsymbol{U}(k-1) \\ &\quad + \sum_{i=0}^{N-1}\boldsymbol{\phi}_L^T(k+i)\boldsymbol{A}^{i}\boldsymbol{B}\Delta \boldsymbol{u}(k) \\ &\quad + \sum_{i=1}^{N-1}\boldsymbol{\phi}_L^T(k+i)\boldsymbol{A}^{i-1}\boldsymbol{B}\Delta \boldsymbol{u}(k+1) \\ &\quad + \sum_{i=2}^{N-1}\boldsymbol{\phi}_L^T(k+i)\boldsymbol{A}^{i-2}\boldsymbol{B}\Delta \boldsymbol{u}(k+2) + \cdots \\ &\quad + \boldsymbol{\phi}_L^T(k+N-1)\boldsymbol{B}\Delta \boldsymbol{u}(k+N-1) \end{aligned} \tag{5}$$

where $N$ refers to the prediction step length, $\Delta \boldsymbol{y}(k+i)$ and $\Delta \boldsymbol{u}(k+i)$ are the incremental form of predictive output and input vector of the system in the future time $k+i$ ($i$=1,2,⋯, $N$ ), respectively. Define several vectors and matrices $\boldsymbol{Y}_N(k)$, $\Delta \boldsymbol{Y}_N(k+1)$, $\Delta \boldsymbol{U}_N(k)$, $\Delta \boldsymbol{U}_{Nu}(k)$, $\bar{\boldsymbol{\Psi}}(k)$ and $\boldsymbol{\Psi}(k)$ by an approximation $\boldsymbol{\phi}_L^T(k+i) = \boldsymbol{\phi}_L^T(k)$ ($i$=1,⋯, $N$-1) as follows and define another set of $\bar{\boldsymbol{\Psi}}(k)$ and $\boldsymbol{\Psi}(k)$ in Appendix II if $\boldsymbol{\phi}_L^T(k+i)$ can be obtained:

$$\Delta \boldsymbol{Y}_N(k+1) = \boldsymbol{Y}_N(k+1) - \boldsymbol{Y}_N(k)$$

$$\boldsymbol{Y}_N(k+1)=\begin{bmatrix}\boldsymbol{y}(k+1)\\ \vdots\\ \boldsymbol{y}(k+N)\end{bmatrix}_{N\bullet My\times 1}\qquad \boldsymbol{E}=\begin{bmatrix}\boldsymbol{I}\\ \vdots\\ \boldsymbol{I}\end{bmatrix}_{N\bullet My\times My}$$

$$\Delta\boldsymbol{U}_N(k)=\begin{bmatrix}\Delta\boldsymbol{u}(k)\\ \vdots\\ \Delta\boldsymbol{u}(k+N-1)\end{bmatrix}_{(N\bullet Mu)\times 1}\qquad \Delta\boldsymbol{U}_{Nu}(k)=\begin{bmatrix}\Delta\boldsymbol{u}(k)\\ \vdots\\ \Delta\boldsymbol{u}(k+N_u-1)\end{bmatrix}_{(Nu\bullet Mu)\times 1}$$

$$\bar{\boldsymbol{\Psi}}(k)=\begin{bmatrix}\boldsymbol{\phi}_L^T(k)\boldsymbol{A}\\ \sum_{i=0}^{1}\boldsymbol{\phi}_L^T(k)\boldsymbol{A}^{i+1}\\ \vdots\\ \sum_{i=0}^{N_u-1}\boldsymbol{\phi}_L^T(k)\boldsymbol{A}^{i+1}\\ \vdots\\ \sum_{i=0}^{N-1}\boldsymbol{\phi}_L^T(k)\boldsymbol{A}^{i+1}\end{bmatrix}=[\bar{\boldsymbol{\Psi}}_1(k),\bar{\boldsymbol{\Psi}}_2(k),\cdots,\bar{\boldsymbol{\Psi}}_{L-1}(k),\boldsymbol{0}]_{(N\bullet My)\times(L\bullet Mu)}$$

where $\bar{\boldsymbol{\Psi}}_j(k)$ denotes the columns from $(j-1)\bullet M_u+1$ to $j\bullet M_u$ in $\bar{\boldsymbol{\Psi}}(k)$.

$$\boldsymbol{\Psi}(k)_{(N\bullet My)\times(N\bullet Mu)}=\begin{bmatrix}\boldsymbol{\phi}_L^T(k)\boldsymbol{B} & \boldsymbol{0} & \cdots & \boldsymbol{0}\\ \sum_{i=0}^{1}\boldsymbol{\phi}_L^T(k)\boldsymbol{A}^i\boldsymbol{B} & \boldsymbol{\phi}_L^T(k)\boldsymbol{B} & \vdots & \vdots\\ \vdots & \vdots & \ddots & \vdots\\ \sum_{i=0}^{N_u-1}\boldsymbol{\phi}_L^T(k)\boldsymbol{A}^i\boldsymbol{B} & \sum_{i=1}^{N_u-1}\boldsymbol{\phi}_L^T(k)\boldsymbol{A}^{i-1}\boldsymbol{B} & \ddots & \boldsymbol{0}\\ \vdots & \vdots & \ddots & \vdots\\ \sum_{i=0}^{N-1}\boldsymbol{\phi}_L^T(k)\boldsymbol{A}^i\boldsymbol{B} & \sum_{i=1}^{N-1}\boldsymbol{\phi}_L^T(k)\boldsymbol{A}^{i-1}\boldsymbol{B} & \cdots & \boldsymbol{\phi}_L^T(k)\boldsymbol{B}\end{bmatrix}$$

Then we rewrite (5) as:

$$\boldsymbol{Y}_N(k+1)=\boldsymbol{E}\boldsymbol{y}(k)+\boldsymbol{\Psi}(k)\Delta\boldsymbol{U}_N(k)+\bar{\boldsymbol{\Psi}}(k)\Delta\boldsymbol{U}(k-1)\tag{6}$$

$N_u$ is referred to the control step length. We consider $\Delta\boldsymbol{u}(k+j-1)=\boldsymbol{0}$, $N_u<j\le N$, (6) is further rewritten as

$$\boldsymbol{Y}_N(k+1)=\boldsymbol{E}\boldsymbol{y}(k)+\boldsymbol{\Psi}_{Nu}(k)\Delta\boldsymbol{U}_{N_u}(k)+\bar{\boldsymbol{\Psi}}(k)\Delta\boldsymbol{U}(k-1)\tag{7}$$

where $\boldsymbol{\Psi}_{Nu}(k)$ is given as

$$\boldsymbol{\Psi}_{Nu}(k)_{(N\bullet My)\times(Nu\bullet Mu)}=\begin{bmatrix}\boldsymbol{\phi}_L^T(k)\boldsymbol{B} & \boldsymbol{0} & \cdots & \boldsymbol{0}\\ \sum_{i=0}^{1}\boldsymbol{\phi}_L^T(k)\boldsymbol{A}^i\boldsymbol{B} & \boldsymbol{\phi}_L^T(k)\boldsymbol{B} & \vdots & \\ \vdots & \vdots & \vdots & \vdots\\ \sum_{i=0}^{Nu-1}\boldsymbol{\phi}_L^T(k)\boldsymbol{A}^i\boldsymbol{B} & \sum_{i=1}^{Nu-1}\boldsymbol{\phi}_L^T(k)\boldsymbol{A}^{i-1}\boldsymbol{B} & \cdots & \boldsymbol{\phi}_L^T(k)\\ \vdots & \vdots & \vdots & \vdots\\ \sum_{i=0}^{N-1}\boldsymbol{\phi}_L^T(k)\boldsymbol{A}^i\boldsymbol{B} & \sum_{i=1}^{N-1}\boldsymbol{\phi}_L^T(k)\boldsymbol{A}^{i-1}\boldsymbol{B} & & \sum_{i=Nu-1}^{N-1}\boldsymbol{\phi}_L^T(k)\boldsymbol{A}^{i-Nu+1}\boldsymbol{B}\end{bmatrix}$$

## III. Model-Free Adaptive Predictive Control Design and Performance Analysis

### A. Design of MFAPC

We choose the following cost function:

$$J=E\{\left[\boldsymbol{Y}_N^*(k+1)-\boldsymbol{Y}_N(k+1)\right]^T\left[\boldsymbol{Y}_N^*(k+1)-\boldsymbol{Y}_N(k+1)\right]+\Delta\boldsymbol{U}_{N_u}^T(k)\boldsymbol{\lambda}\Delta\boldsymbol{U}_{N_u}(k)\}\tag{8}$$

where $\boldsymbol{\lambda}$=diag($\lambda_1,\cdots,\lambda_{N\times Mu}$) is the weighted diagonal matrix whose elements can be positive or negative, though it is likely that seldom literature uses the negative values; $[\tilde{\boldsymbol{Y}}_N^*(k+1)]^T=\left[[\boldsymbol{y}^*(k+1)]^T,\cdots,[\boldsymbol{y}^*(k+N)]^T\right]$ is composed of $\boldsymbol{y}^*(k+i)$ ($i$=1,2,···, $N$) which is the desired output vector of the system in future time $k+i$.

In the same way as [34]-[36], we substitute (7) into (8) and find the controller by solving the optimal solution of (8) $\dfrac{\partial J}{\partial\Delta\boldsymbol{U}_{Nu}(k)}=\boldsymbol{0}$. Then we may originate the controller into

$$\Delta\boldsymbol{U}_{Nu}(k)=[\boldsymbol{\Psi}_{Nu}^T(k)\boldsymbol{\Psi}_{Nu}(k)+\boldsymbol{\lambda}]^{-1}\boldsymbol{\Psi}_{Nu}^T(k)[\boldsymbol{Y}_N^*(k+1)-\boldsymbol{E}\boldsymbol{y}(k)-\bar{\boldsymbol{\Psi}}(k)\Delta\boldsymbol{U}(k-1)]\tag{9}$$

Then the current input is provided by

$$\boldsymbol{u}(k)=\boldsymbol{u}(k-1)+\boldsymbol{g}^T\Delta\boldsymbol{U}_{Nu}(k)\tag{10}$$

where $\boldsymbol{g}=\left[\boldsymbol{I},\boldsymbol{0},\cdots,\boldsymbol{0}\right]^T$.

### B. Stability Analysis

This part provides the performance analysis of MFAPC.

Define

$$\boldsymbol{\phi}_L(z^{-1})=\boldsymbol{\Phi}_1(k)+\cdots+\boldsymbol{\Phi}_L(k)z^{-L+1}\tag{11}$$

$$\boldsymbol{P}(k)=[\boldsymbol{\Psi}_{Nu}^T(k)\boldsymbol{\Psi}_{Nu}(k)+\boldsymbol{\lambda}]^{-1}\boldsymbol{\Psi}_{Nu}^T(k)\tag{12}$$

where $z^{-1}$ denotes backward shift operator, and $\Delta=1-z^{-1}$.

We can rewrite (3) as

$$\Delta\boldsymbol{y}(k+1)=\boldsymbol{\phi}_L(z^{-1})\Delta\boldsymbol{u}(k)\tag{13}$$

We combine (9), (10) and (13) to obtain

$$\left[\Delta\boldsymbol{I}+z^{-1}\boldsymbol{\phi}_L(z^{-1})(\boldsymbol{I}+z^{-1}\boldsymbol{P}(k)\bar{\boldsymbol{\Psi}}(k)\boldsymbol{T}_u)^{-1}\boldsymbol{P}(k)\boldsymbol{E}\right]\boldsymbol{y}(k+1)=\boldsymbol{\phi}_L(z^{-1})(\boldsymbol{I}+z^{-1}\boldsymbol{P}(k)\bar{\boldsymbol{\Psi}}(k)\boldsymbol{T}_u)^{-1}\boldsymbol{P}(k)\boldsymbol{H}\boldsymbol{y}^*(k+1)\tag{14}$$

where $\boldsymbol{H}=[\boldsymbol{I},z\boldsymbol{I},\cdots,z^{N-1}\boldsymbol{I}]^T$ and $\boldsymbol{T}_u=[\boldsymbol{I},z^{-1}\boldsymbol{I},\cdots,z^{-Lu+1}\boldsymbol{I}]^T$.

If rank$\left[\boldsymbol{\phi}_L(z^{-1})\right]=M_y$ ($M_u\ge M_y$), by choosing $N$, $N_u$ and $\boldsymbol{\lambda}$, we may attain the following inequality:

$$\boldsymbol{T}=\Delta\boldsymbol{I}+z^{-1}\boldsymbol{\phi}_L(z^{-1})(\boldsymbol{I}+z^{-1}\boldsymbol{P}(k)\bar{\boldsymbol{\Psi}}(k)\boldsymbol{T}_u)^{-1}\boldsymbol{P}(k)\boldsymbol{E},\qquad |z|>1\tag{15}$$

for the stability of the system.

Further, if the system stability is guaranteed, sometimes we may determine the steady-state error vector by

$$\lim_{k\to\infty}\boldsymbol{e}(k)=\lim_{z\to 1}(1-z^{-1})\boldsymbol{T}^{-1}(z^{-1})\left[\Delta\boldsymbol{I}-\boldsymbol{\phi}_L(z^{-1})\bullet(\boldsymbol{I}+z^{-1}\boldsymbol{P}(k)\tilde{\boldsymbol{\Psi}}_U(k)\boldsymbol{T}_u)^{-1}\boldsymbol{P}(k)(\boldsymbol{H}-z^{-1}\boldsymbol{E})\right]\boldsymbol{Z}(\boldsymbol{y}^*(k+1))\tag{16}$$

When the desired trajectories are step signals $\boldsymbol{y}^*(k)=[1,\cdots,1]_{1\times My}^T$ ($i$=1, ···, $N$), we may remove the steady-state error by choosing $\boldsymbol{\lambda}=\boldsymbol{0}$.

### *C. Estimate Method*

The adaptability to nonlinearity or uncertainty would be achieved by combining one online parameter estimation with the control scheme according to the certainty equivalence adaptive control [25], [26]. In this paper, we make an attempt to replace the current projection algorithm or least-square method with the BPNN and RBFNN algorithms.

*1)* BPNN estimation algorithm

Without loss of generality, a feedforward neural network with a single hidden layer constituted by $m$ nodes is applied in this part. The regressor in the input layer is constituted by

$$\boldsymbol{O}^{(1)}(k) = [\boldsymbol{u}^T(k-1), \cdots, \boldsymbol{u}^T(k-L)]^T \tag{17}$$

The output vector of the network is

$$\boldsymbol{y}^N(k) = [y_1^N(k), \cdots, y_{My}^N(k)]^T \tag{18}$$

The $t$-th output of the network is represented by $y_t^N(k)$, ($t$=1, ···, $M_y$). The activation function of hidden layer is normally taken with sigmoid function

$$\zeta(\boldsymbol{x}) = \frac{1}{1+e^{-\boldsymbol{x}}} \tag{19}$$

The induced local field vector, which acts as the input of the activation function in the hidden layer, is defined by

$$net_t^{(2)}(k) = \boldsymbol{w}_t^{(1)}(k)\boldsymbol{O}^{(1)}(k) \tag{20}$$

where $\boldsymbol{w}_t^{(1)}(k)$ is the synaptic weight matrix between the input and hidden layers. $\boldsymbol{O}_t^{(2)}(k)$ denotes the output of the hidden layer at time $k$ and is computed by

$$\boldsymbol{O}_t^{(2)}(k) = \zeta\left(net_t^{(2)}(k)\right) \tag{21}$$

We choose the index function as

$$E_t(k) = \frac{1}{2}(y_t(k) - y_t^N(k))^2 \tag{22}$$

The $t$-th output of the network is computed by

$$y_t^N(k) = \boldsymbol{w}_t^{(2)}(k)\boldsymbol{O}_t^{(2)}(k) \tag{23}$$

where $\boldsymbol{w}_t^{(2)}(k)$ is the synaptic weight vector from the hidden neurons to output neurons. Differentiating the both sides of (23) with respect to $\boldsymbol{w}_t^{(1)}(k)$ and $\boldsymbol{w}_t^{(2)}(k)$ yields

$$\frac{\partial E_t(k)}{\partial \boldsymbol{w}_t^{(2)}(k)} = -e_t(k)\boldsymbol{O}_t^{(2)}(k) \tag{24}$$

$$\frac{\partial E_t(k)}{\partial \boldsymbol{w}_t^{(1)}(k)} = -e_t(k)diag(\boldsymbol{w}_t^{(2)}(k))\zeta'\left(net_t^{(2)}(k)\right)\left[\boldsymbol{O}^{(1)}(k)\right]^T \tag{25}$$

$$e_t(k) = y_t(k) - y_t^N(k) \tag{26}$$

where $diag(\bullet)$ represents the matrix operation which converts the vector into a diagonal matrix. Then the correction $\Delta\boldsymbol{w}_t^{(l)}(k)$ applied to $\boldsymbol{w}_t^{(l)}(k)$ is calculated by using the method of gradient descent:

$$\Delta\boldsymbol{w}_t^{(l)}(k) = -\eta\frac{\partial E_t(k)}{\partial \boldsymbol{w}_t^{(l)}(k)} \quad (l=1, 2) \tag{27}$$

where $\eta$ is the learning-rate parameter of the back-propagation algorithm.

The synaptic weight $\boldsymbol{w}_t^{(l)}(k)$ is adjusted by the generalized delta rule

$$\boldsymbol{w}_t^{(l)}(k) = \boldsymbol{w}_t^{(l)}(k-1) + \Delta\boldsymbol{w}_t^{(l)}(k) + \alpha(\boldsymbol{w}_t^{(l)}(k-1) - \boldsymbol{w}_t^{(l)}(k-2)) \tag{28}$$

and $\alpha$ is the momentum constant.

Then we obtain the online estimation parameters for the $t$-th row of PJM

$$\boldsymbol{K}_t\boldsymbol{\phi}_L^T(k) = \boldsymbol{K}_t\left[\frac{\partial \boldsymbol{f}(\boldsymbol{\varphi}(k-1))}{\partial \boldsymbol{u}^T(k-1)}, \cdots, \frac{\partial \boldsymbol{f}(\boldsymbol{\varphi}(k-1))}{\partial \boldsymbol{u}^T(k-L)}\right] = \left[\frac{\partial y_t^N(k)}{\partial \boldsymbol{O}^{(1)}(k)}\right]^T = \left[diag(\boldsymbol{w}_t^{(2)}(k))\zeta'\left(net_t^{(2)}(k)\right)\right]^T \boldsymbol{w}_t^{(1)}(k) \tag{29}$$

where $\boldsymbol{K}_t = \begin{bmatrix} 0_1 & \cdots & 0_{t-1} & 1 & 0_{t+1} & \cdots & 0 \end{bmatrix}_{1\times My}$.

In order to improve the precision of the estimation method, we may obtain the offline estimated synaptic weights by restricting the bound of the following index function

$$E_t(k) = \sum_{k=1}^{num}\frac{1}{2}e_t^2(k) \tag{30}$$

where $num$ is the number of training epochs.

Similarly, we can use the BPNN with $n$ hidden layers to estimate $\boldsymbol{\phi}_L^T(k) = \left[\frac{\partial \boldsymbol{f}(\boldsymbol{\varphi}(k-1))}{\partial \boldsymbol{u}^T(k-1)}, \cdots, \frac{\partial \boldsymbol{f}(\boldsymbol{\varphi}(k-1))}{\partial \boldsymbol{u}^T(k-L)}\right]$ through (33), according to the chain rule of partial derivatives. The corresponding architectural graph of neural networks with $n$ hidden layers is shown in Fig. 1.

$$\boldsymbol{K}_t\boldsymbol{\phi}_L^T(k) = \boldsymbol{K}_t\left[\frac{\partial \boldsymbol{f}(\boldsymbol{\varphi}(k-1))}{\partial \boldsymbol{u}^T(k-1)}, \cdots, \frac{\partial \boldsymbol{f}(\boldsymbol{\varphi}(k-1))}{\partial \boldsymbol{u}^T(k-L)}\right] = \left[\frac{\partial y_t^N(k)}{\partial \boldsymbol{O}^{(1)}(k)}\right]^T = \left[\cdots\left[diag\left(\left[diag(\boldsymbol{w}_t^{(n+1)}(k))\zeta'\left(net_t^{(n+1)}(k)\right)\right]^T \boldsymbol{w}_t^{(n)}(k)\right)\zeta'\left(net_t^{(n)}(k)\right)\right]\cdots\zeta'\left(net_t^{(2)}(k)\right)\right]^T \boldsymbol{w}_t^{(1)}(k) \tag{31}$$

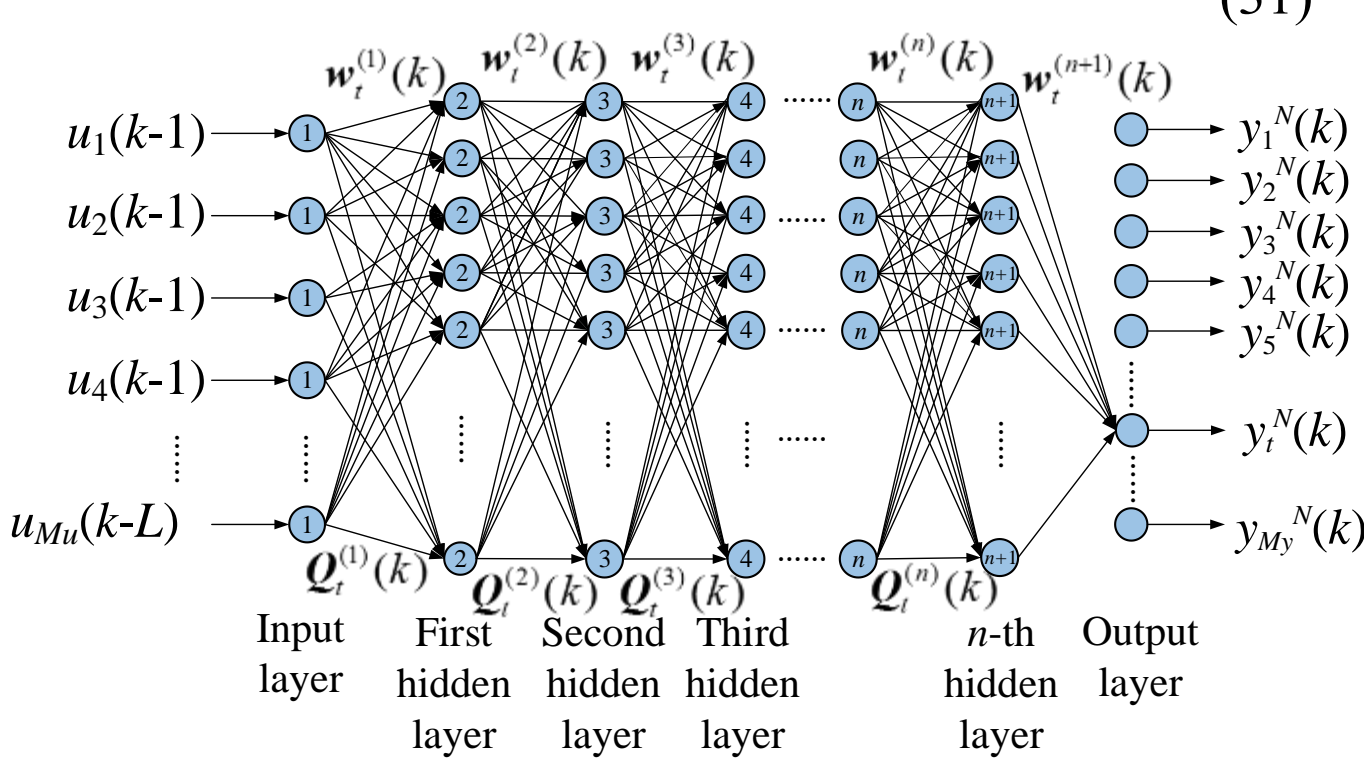

Fig. 1 Neural Networks with $n$ hidden layers, an input layer of $M_u \times L$ nodes and an output layer of $M_y$ nodes

For example, when the number of hidden layers is $n$=2, we have

$$\boldsymbol{K}_t\boldsymbol{\phi}_L^T(k) = \left[diag\left(\left[diag(\boldsymbol{w}_t^{(3)}(k))\zeta'\left(net_t^{(3)}(k)\right)\right]^T \boldsymbol{w}_t^{(2)}(k)\right)\zeta'\left(net_t^{(2)}(k)\right)\right]^T \bullet\boldsymbol{w}_t^{(1)}(k) \tag{32}$$

*2)* RBFNN estimation algorithm

Similar to the BPNN with one hidden layer, the activation function of the hidden layer is chosen with the Gaussian radial-basis-function

$$R_{jt}(\boldsymbol{O}^{(1)}(k)) = e^{-\frac{\left\|\boldsymbol{O}^{(1)}(k)-\boldsymbol{c}_{jt}\right\|^2}{2b_{jt}^2}}, j=1, \cdots, m;\ t=1, \cdots, M_y \quad (33)$$

where $c_{jt}$ and $b_{jt}$ represent the center vector and the radius of the $j$-th hidden node, respectively.

And we construct

$$\boldsymbol{R}_t(\boldsymbol{O}^{(1)}(k)) = [R_{1t}(\boldsymbol{O}^{(1)}(k)),\cdots,R_{mt}(\boldsymbol{O}^{(1)}(k))]^T \quad (34)$$

The synaptic weight vector between the input layer and the hidden layer is expressed by $\boldsymbol{w}^{(1)}(k)=[1,\cdots,1]$; The synaptic weight vector between the hidden layer and the $t$-th output node $\boldsymbol{w}_t^{(2)}(k)$ is the same as that of above BPNN. In the same process, we can obtain the online estimate parameters for the $t$-th row of PJM

$$\begin{aligned}\boldsymbol{K}_t\boldsymbol{\phi}_L^T(k) &= \left[\frac{\partial y_t^N(k)}{\partial \boldsymbol{O}^{(1)}(k)}\right]^T \\ &= \left[\boldsymbol{\Gamma}\left(\left[diag(\boldsymbol{w}_t^{(2)}(k))\boldsymbol{R}_t(\boldsymbol{O}^{(1)}(k))\right]^T, L\times M_u\right)\boldsymbol{M}(k)\right]^T\end{aligned} \quad (35)$$

where $\boldsymbol{M}(k)=\left[\left(\frac{\boldsymbol{O}^{(1)}(k)-\boldsymbol{c}_{1t}}{2b_{1t}^2}\right)^T,\cdots,\left(\frac{\boldsymbol{O}^{(1)}(k)-\boldsymbol{c}_{mt}}{2b_{mt}^2}\right)^T\right]^T$ and $\boldsymbol{\Gamma}(\boldsymbol{v},p)$ represents the matrix operation which transforms the vector $\boldsymbol{v}=[v_1,\cdots,v_m]$ into the matrix $\boldsymbol{\Gamma}(\boldsymbol{v},p)=[v_1\boldsymbol{I}_{p\times p},\cdots,v_m\boldsymbol{I}_{p\times p}]$. One can refer to the textbook [37] about neural network for the meaning and iteration scheme of parameters $\boldsymbol{c}_{jt}$, $b_{jt}$ and $\boldsymbol{w}_t^{(2)}(k)$.

*Remark 1*: The above estimate methods can be applied in the full-form MFAPC in [21] and full-form MFAC in [22], according to the chain rule of partial derivatives. Take the above BPNN as an example. The regressor will be

$$\boldsymbol{O}^{(1)}(k)=[\boldsymbol{y}^T(k-1),\cdots,\boldsymbol{y}^T(k-L_y),\boldsymbol{u}^T(k-1),\cdots,\boldsymbol{u}^T(k-L_u)]^T \quad (36)$$

Similarly, the corresponding PJM in [21], [22] is described by

$$\begin{aligned}\boldsymbol{K}_t\boldsymbol{\phi}_L^T(k) &= \left[\frac{\partial y_t^N(k)}{\partial \boldsymbol{O}^{(1)}(k)}\right]^T = \\ &\left[\cdots\left[diag\left(\left[diag(\boldsymbol{w}_t^{(n+1)}(k))\zeta'\left(net_t^{(n+1)}(k)\right)\right]^T\boldsymbol{w}_t^{(n)}(k)\right)\zeta'\left(net_t^{(n)}(k)\right)\right]^T\right. \\ &\left.\cdots\zeta'\left(net_t^{(2)}(k)\right)\right]^T\boldsymbol{w}_t^{(1)}(k)\end{aligned} \quad (37)$$

*3)* Neural network for tuning parameters in controller

We consider that the regressor is constituted by

$$\boldsymbol{O}^{(1)}(k)=[\lambda_1(k),\cdots,\lambda_{Nu\times Mu}(k)]^T \quad (38)$$

The index function is chosen with

$$E(k)=\frac{1}{2}\sum_{t=1}^{My}l_t(y_t(k)-y_t^N(k))^2 \quad (39)$$

where $l_t$ represents the weight constant. In a similar way to the aforementioned BPNN or RBFNN iteration, we can obtain the correction of target parameters

$$\Delta\boldsymbol{O}^{(1)}(k)=-\eta\frac{\partial E(k)}{\partial \boldsymbol{O}^{(1)}(k)} \quad (40)$$

which renders the adjusting $\boldsymbol{\lambda}$ method (43).

$$\boldsymbol{O}^{(1)}(k+1)=\boldsymbol{O}^{(1)}(k)+\Delta\boldsymbol{O}^{(1)}(k)+\alpha(\boldsymbol{O}^{(1)}(k-1)-\boldsymbol{O}^{(1)}(k-2)) \quad (41)$$

## IV. SIMULATIONS

Two simple yet nontrivial examples are carried out to illustrate the feasibility and effectiveness of the proposed MFAPC with NNs just serving as its parameter estimator. The third example shows an interesting finding about the adjustment of elements in matrix $\boldsymbol{\lambda}$ by BPNN.

Example 1: In this example, we use the aforementioned BPNN estimate method to design the MFAPC controller. The model is given as the following system:

$$y(k+1)=u_1(k)+0.4u_2^3(k)+0.5u_1(k-1)+0.6u_2(k-1) \quad (42)$$

The desired output trajectory is

$$y_d(k+1)=\sin(\pi k/100) \quad (43)$$

The initial values are $y(1)=y(2)=0$, $\boldsymbol{u}(1)=\boldsymbol{u}(2)=\boldsymbol{0}$. The structure parameters of the controller are $L=2$, $M_y=1$ and $M_u=2$. We choose $\boldsymbol{\lambda}_{MFAPC}=0.01\boldsymbol{I}$, $N=2$ and $N_u=2$.

The estimation method is performed through BPNN which has 4 input nodes, 1 output node and a single hidden layer constituted of 6 neurons. The regressor is taken with $\boldsymbol{O}^{(1)}(k)=[u_1(k),u_2(k),u_1(k-1),u_2(k-1)]$. The initial settings of the synaptic weights matrices $\boldsymbol{w}^{(1)}$ and $\boldsymbol{w}^{(2)}$ are randomly chosen in the interval [-1,1].

$$[w^{(1)}]^T=0.1\bullet\begin{bmatrix}-8 & -1 & -6 & -7 & 9 & 1\\ 4 & 5 & -9 & 3 & 4 & -2\\ 1 & -5 & 4 & 6 & -2 & 7\\ 7 & -1 & -3 & 2 & 8 & -3\end{bmatrix} \quad (44)$$

$$w^{(2)}=0.1\bullet\begin{bmatrix}-1 & -3 & 7 & -5 & 8 & -2\end{bmatrix} \quad (45)$$

The parameters of learning iteration are chosen with $\eta=0.5$ and $\alpha=0.05$. The training and test data are generated by inputting the system (42) with the following controller outputs:

$$\begin{aligned}u_1(k)&=0.9*\sin(\pi k/100)\quad (k=1:900)\\ u_2(k)&=0.6*\sin(\pi k/100)\end{aligned} \quad (46)$$

We choose the index function (30) with such a stopping criterion: $\sum_{k=1}^{num}\frac{1}{2}e^2(k)<0.002$, whose corresponding number of epochs of training is $num=7453$.

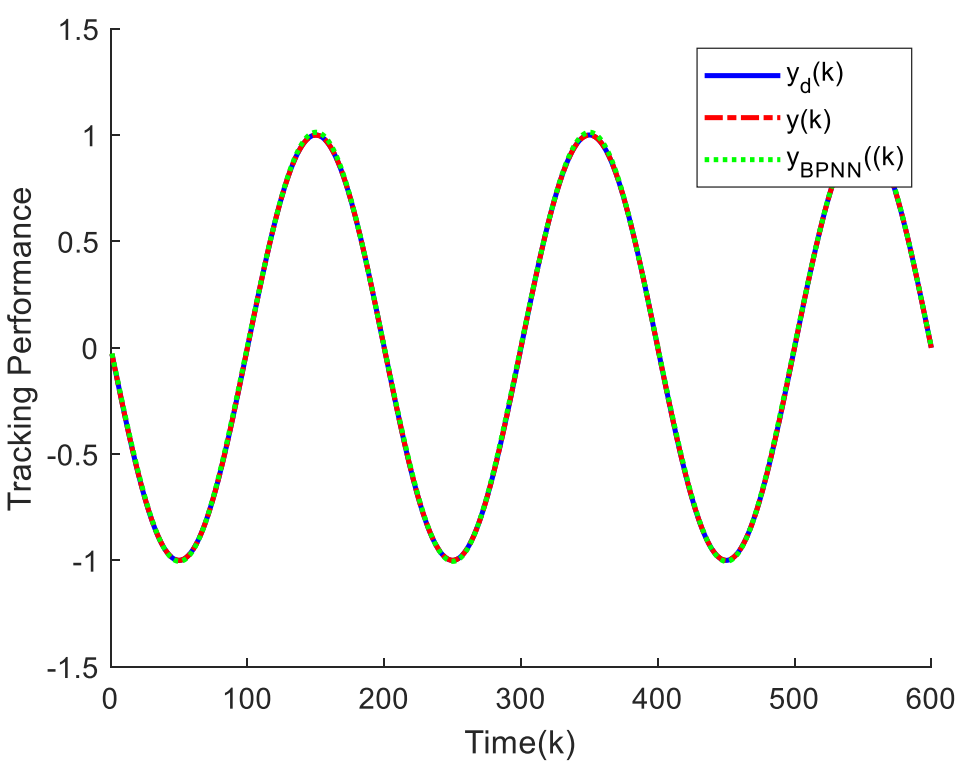

Fig. 2 Tracking performance

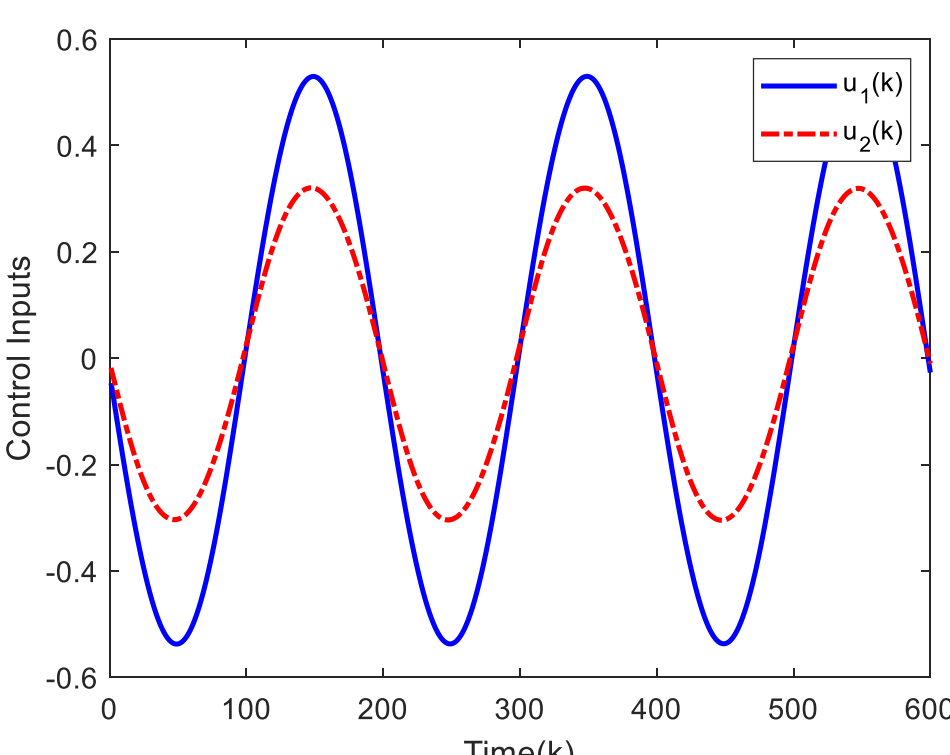

Fig. 3 Control inputs

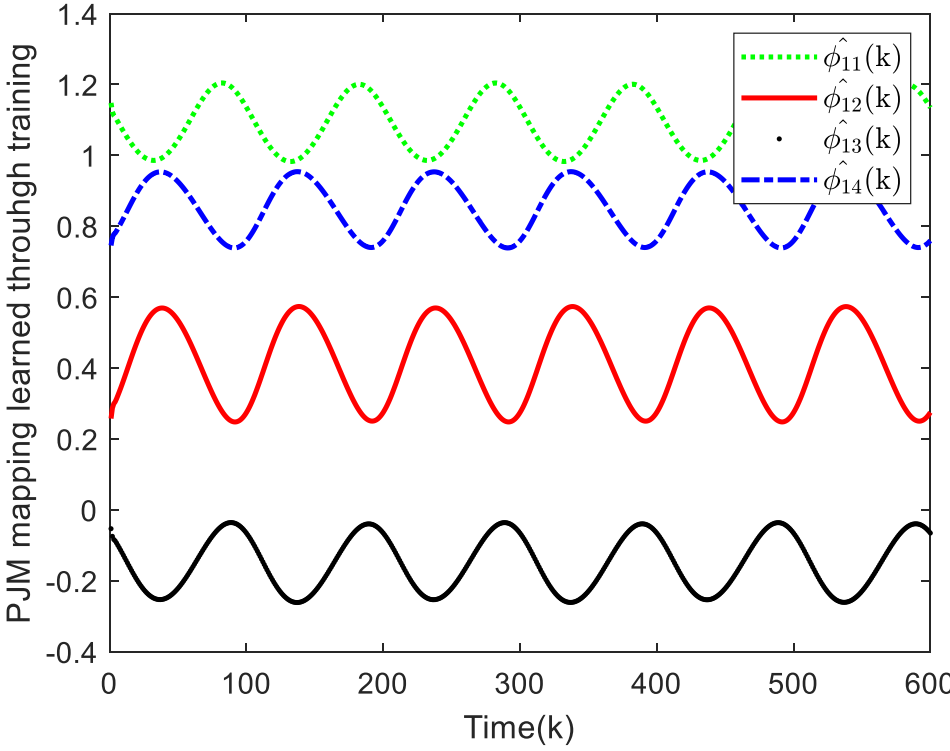

Fig. 4 Elements in estimated PJM

The output $y(k)$ of the system controlled by the proposed MFAPC and the simultaneous output $y_{BPNN}(k)$ of the network driven by the MFAPC outputs are depicted in Fig. 2. The outputs of the controller and the elements in PJM mapping learned through training are shown in Fig. 3 and Fig. 4, respectively. Fig. 2 shows that the output of the system tracks the desired output with high performance, and the network can reproduce the correct input and output mapping of the system, although, in Fig. 4, it is observed that the estimated parameters do not converge even after extensive off-line training.

Example 2: In this simulation, we use the online RBFNN to build the model and to calculate the MFAPC controller outputs.

The parameters in the RBFNN estimate method are chosen with $\eta = 0.5$ and $\alpha = 0.05$ . The index function is applied with (22).

The desired output trajectory is

$$y_d(k) = 2 + \sin(\pi k / 100) \tag{47}$$

The structure of the adopted RBFNN is identical to that in Example 1. The initial values are $b_j = 1$ , $\boldsymbol{c}_j = 0.01\bullet[1,1,1,1]^T$ ($j$=1,···,6) and $\boldsymbol{w}^{(2)} = rands(1,6)$ .

The output of the system controlled by the proposed MFAPC and the network output $y_{RBFNN}(k)$ are shown in Fig. 5. It is evident that the generalization of online RBFNN is not as good as that of the former batch learning of BPNN. The merits of BPNN lie in its accurate estimation and parallelization of the learning process.

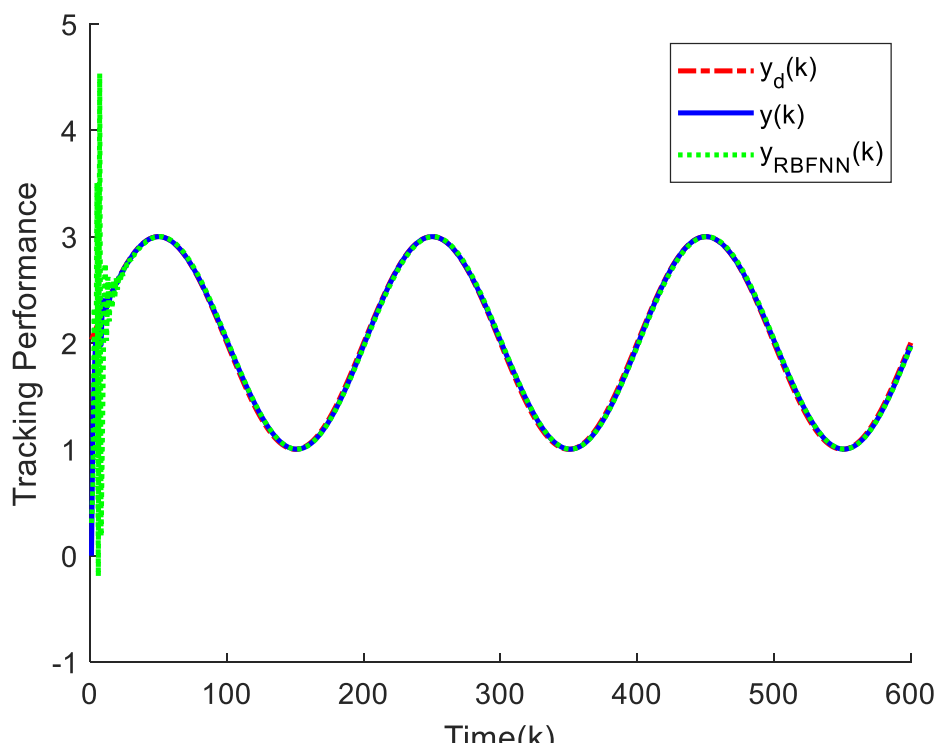

Fig. 5 Tracking performance

Example 3: The model is formulated as:

$$\begin{aligned} y_1(k+1) &= u_1(k) + 0.4u_2(k) + 0.5u_1(k-1) + 0.6u_2(k-1) \\ y_2(k+1) &= 0.8u_1(k) + 1.2u_2(k) + 0.4u_1(k-1) + 0.7u_2(k-1) \end{aligned} \tag{48}$$

The desired output trajectories are

$$\begin{aligned} y_{d1}(k) &= \sin(\pi k / 100) \\ y_{d2}(k) &= \sin(\pi k / 10) \end{aligned} \tag{49}$$

The controller coefficients are consistent with the real coefficient matrices. The structure of online BPNN, learning-rate parameters and momentum constants for synaptic weights are chosen to be the same as those in Example 1. The input of network is $\boldsymbol{O}^{(1)}(k) = [\lambda_1(k), \cdots, \lambda_{Nu\times Mu}(k)]^T$ which is initialized with a zero vector. The final learning rate parameter and momentum constant for the iteration of $\boldsymbol{O}^{(1)}(k)$ are assigned $\eta = 0.5$ and $\alpha = 0$ , respectively. The initializations for synaptic weight matrices are chosen as follows:

$$[\boldsymbol{w}^{(1)}]^T = 0.1\bullet\begin{bmatrix} -1 & -1 & -1 & -1 & -1 & -1 \\ 2 & 2 & 2 & 2 & 2 & 2 \\ -3 & -3 & -3 & -3 & -3 & -3 \\ 4 & 4 & 4 & 4 & 4 & 4 \end{bmatrix} \tag{50}$$

$$\boldsymbol{w}^{(2)} = 0.1\bullet\begin{bmatrix} 1 & -2 & 3 & -4 & 5 & -6 \end{bmatrix} \tag{51}$$

Fig. 6 and Fig. 7 show the outputs of the system. Fig. 8 shows the control outputs. Fig. 9 shows the elements in $\boldsymbol{\lambda}$.

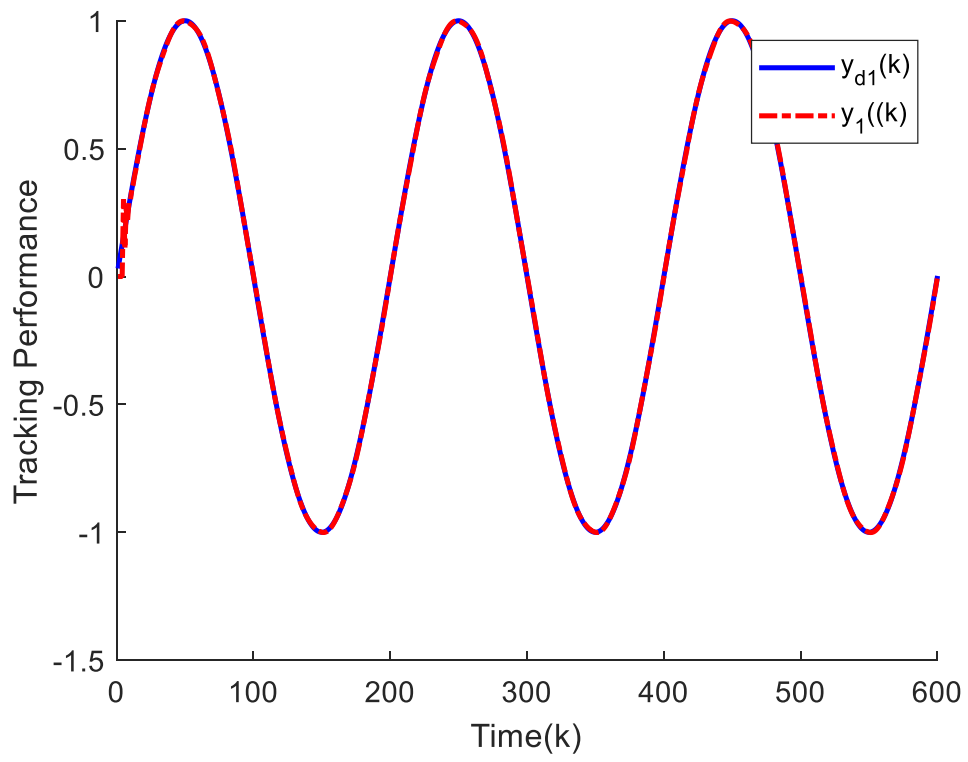

Fig. 6 Tracking performance of $y_1$

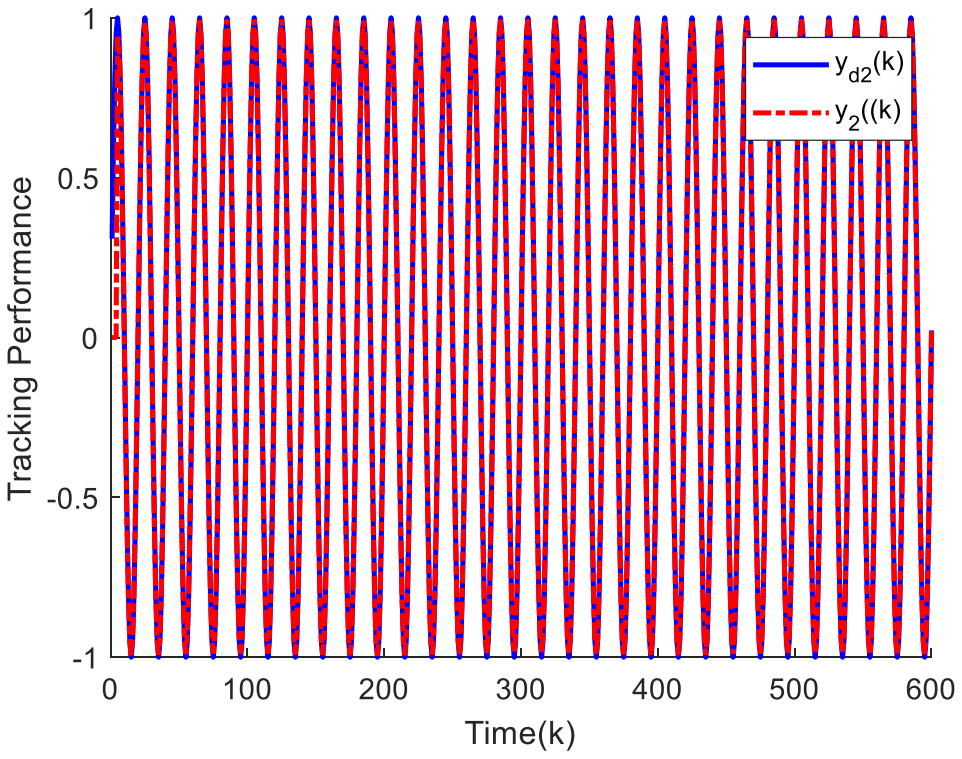

Fig. 7 Tracking performance of $y2$

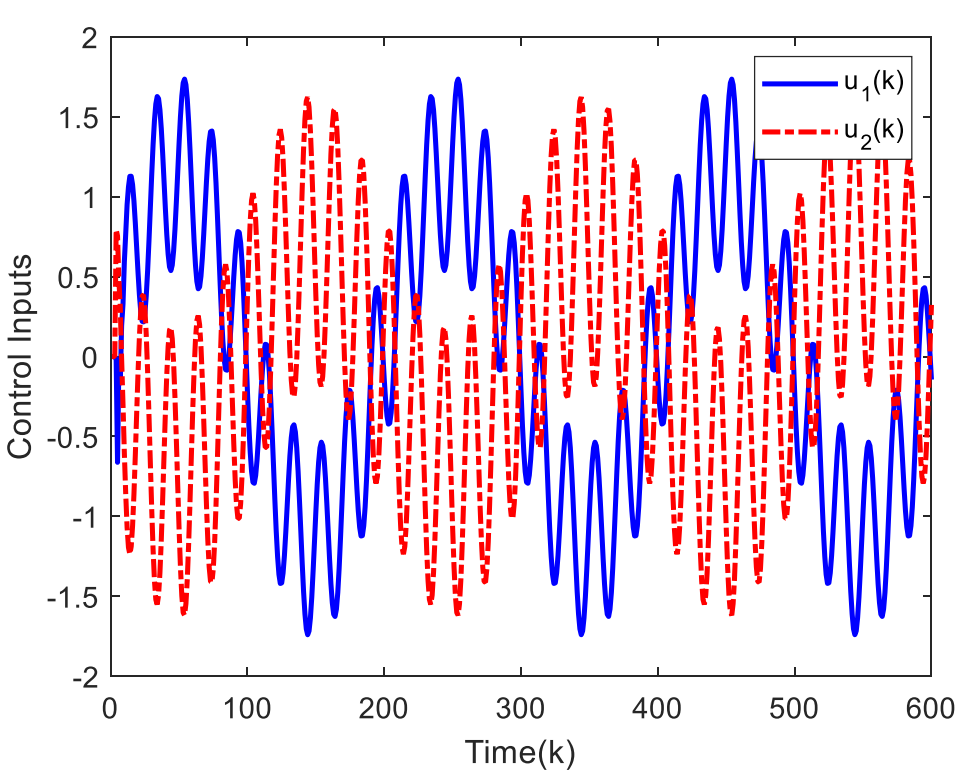

Fig. 8 Control inputs

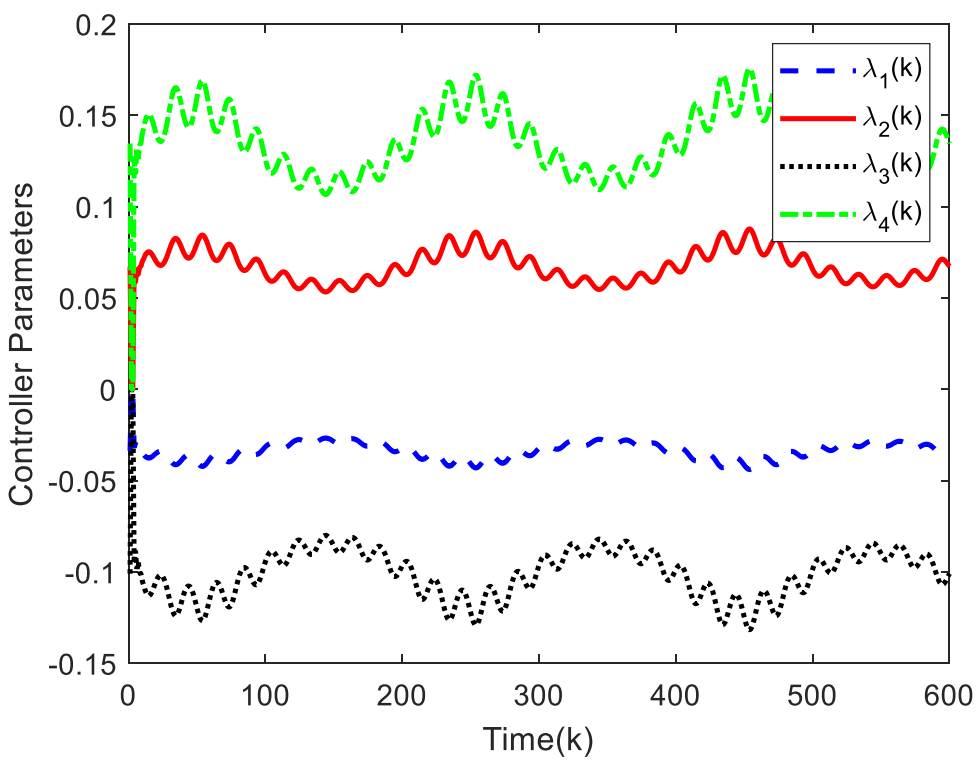

Fig. 9 Elements in $\boldsymbol{\lambda}$

It is quite straightforward to observe that two elements in matrix $\boldsymbol{\lambda}$ shown in Fig. 9 are negative, whereas they are normally defined as the positive weights of index function in the literature. More interestingly, many simulations indicate that the control performance may be guaranteed as well when using negative elements in the weight matrix $\boldsymbol{\lambda}$, and the rationale can be explained by (15). Furthermore, we can replace $\boldsymbol{\lambda}$ with $\boldsymbol{\lambda}(1-z^{-1})$ to introduce another integrator, which reduces the static error of ramp-response while meeting the requirement of $\boldsymbol{\lambda} \neq \mathbf{0}$ and simultaneously mitigating the influence caused by the disturbance $\boldsymbol{d} = [a_1,\cdots,a_{My}]^T \bullet k$ in which $a_i$ ($i$=1,⋯$M_y$) is the constant.

*Remark 2*: We aim to show that the objective of MFAPC and current MFAC designed via partial-form of EDLM is not presupposed in unstable systems. It is illustrated as follows.

The corrected full-form EDLM in [21] and [22] is described as

$$[\boldsymbol{I}-z^{-1}\boldsymbol{\phi}_{Ly}(z^{-1})]\Delta\boldsymbol{y}(k+1)=\boldsymbol{\phi}_{Lu}(z^{-1})\Delta\boldsymbol{u}(k) \tag{52}$$

Due to $[\boldsymbol{I}-z^{-1}\boldsymbol{\phi}_{Ly}(z^{-1})]$ is invertible, (54) can be rewritten as

$$\Delta\boldsymbol{y}(k+1)=\boldsymbol{\phi}_{L}(z^{-1})\Delta\boldsymbol{u}(k) \triangleq [\boldsymbol{I}-z^{-1}\boldsymbol{\phi}_{Ly}(z^{-1})]^{-1}\boldsymbol{\phi}_{Lu}(z^{-1})\Delta\boldsymbol{u}(k) \tag{53}$$

Herein, one can easily design the partial-form of MFAC controller for the linear system (52) as long as $\boldsymbol{I}-z^{-1}\boldsymbol{\phi}_{Ly}(z^{-1})$ is stable.

Now, a simple instance is given to show its deficiency when $\boldsymbol{I}-z^{-1}\boldsymbol{\phi}_{Ly}(z^{-1})$ is unstable.

Consider the SISO linear system

$$y(k+1)=1.1\bullet y(k)+u(k) \tag{54}$$

From (54), it is easy for someone to have the finite impulse response polynomial:

$$\boldsymbol{\phi}_L(z^{-1})=1+1.1z^{-1}+\left(1.1\right)^2 z^{-2}+\cdots+\phi_i z^{-i}+\cdots \tag{55}$$

where $|\phi_i|=\left(1.1\right)^{i-1}<|\phi_{i+1}|$.

In this instance, one may attempt to design the partial-form of MFAC but may encounter difficulties in maintaining system stability. Typically, the partial form of MFAC is used in conservatively stable systems, while the full-form of MFAC offers significant advantages in unstable systems. Only when we face their inadequacies can they exert their superiorities in a correct manner.

*Remark 3*: One inherent advantage of the MFAPC is that the $\boldsymbol{P}^T(\boldsymbol{Y}_N^*(k+1)-\boldsymbol{E}\boldsymbol{y}(k))$ acting as the integrator is separated. When the controller design is improper due to inaccurate modeling, or some system behaviors need to be further improved, we may replace $\boldsymbol{P}^T(\boldsymbol{Y}_N^*(k+1)-\boldsymbol{E}\boldsymbol{y}(k))$ by

$$\begin{aligned}&\boldsymbol{P}^T\Big\{\boldsymbol{K}_I(\boldsymbol{Y}_N^*(k+1)-\boldsymbol{E}\boldsymbol{y}(k))+\\&\quad\boldsymbol{K}_P\left[(\boldsymbol{Y}_N^*(k+1)-\boldsymbol{E}\boldsymbol{y}(k))-(\boldsymbol{Y}_N^*(k)-\boldsymbol{E}\boldsymbol{y}(k-1))\right]\Big\}\end{aligned} \tag{56}$$

The parameter matrix $\boldsymbol{K}_P$ and $\boldsymbol{K}_I$ are introduced to change the system behaviors according to PID tuning experience. This is generally viewed as the last procedure or last resort.

The above perspectives are the lessons from the practical experiments.

By the way, the PID tuning experience in *Remark 2* may be replaced with typical behaviors such as tuning the PID using empirical fuzzy control, adaptive control and training methods.

*Remark 4*: If the system is strongly nonlinear, the obtained $\boldsymbol{\phi}_L^T(k)$, $\bar{\boldsymbol{\Psi}}(k)$ and $\boldsymbol{\Psi}(k)$ normally vary evidently from time $k$ to $k$+1 and usually result in poor system performance. Therefore, we recommend using iterative MFAPC or MFAC controller in [21], [22]. The controller is

$$\Delta\boldsymbol{u}(k,i)=\boldsymbol{g}^T[\boldsymbol{\Psi}^T(k,i)\boldsymbol{\Psi}(k,i)+\lambda(k,i)]^{-1}\boldsymbol{\Psi}^T(k,i)[\boldsymbol{Y}_N^*(k+1) \\ -\boldsymbol{E}\boldsymbol{y}(k,i)-\bar{\boldsymbol{\Psi}}(k,i)\Delta\boldsymbol{U}(k-1,i)] \tag{57}$$

where $\bar{\boldsymbol{\Psi}}(k,i)$, $\boldsymbol{\Psi}(k,i)$ and $\Delta\boldsymbol{U}(k-1,i)$ are listed in Appendix II and we choose

$$\boldsymbol{\phi}_L^T(k,i)=[\frac{\partial\boldsymbol{f}(\boldsymbol{\varphi}(k-1))}{\partial\boldsymbol{u}^T(k-1,i)},\cdots,\frac{\partial\boldsymbol{f}(\boldsymbol{\varphi}(k-1))}{\partial\boldsymbol{u}^T(k-n_u-1,i)}] \tag{58}$$

where $\boldsymbol{\phi}_L^T(k,i)$ is online calculated by substituting the system inputs and outputs into the gradient of the system function; $i$ denotes the iteration number before the control inputs are sent to the system at the time of $k$+1. Please refer to [22] and [23] for the applications of the iterative MFAPC.

## V. Conclusion

Based on a class of prediction model that stems from the corrected partial-form of EDLM, a MIMO-MFAPC method is proposed. This method specifically addresses the time-delay problem in multivariable systems and has a wider range of applications compared to the current MFAC. Furthermore, we proved the stability of the system through the performance analysis. Then the NN is introduced for parameter estimation of the controller. Simulations demonstrate the feasibility of this method and showcase remarkable results.

## VI. Appendix I

Proof of Theorem 1

*Proof*: According to [21] we can transform (1) into (52) and then rewrite (52) into (53). Then we finish the proof of Theorem 1 for (1).

Herein, we will prove the Theorem 1 for (2).

Case 1: The ideal pseudo order is $L=n_u+1$ for the EDLM.

On the basis of *Assumption 1* and the definition of differentiability in [38], (2) becomes

$$\Delta\boldsymbol{y}(k+1)=\frac{\partial\boldsymbol{f}(\boldsymbol{\varphi}(k-1))}{\partial\boldsymbol{u}^T(k-1)}\Delta\boldsymbol{u}(k)+\cdots+\frac{\partial\boldsymbol{f}(\boldsymbol{\varphi}(k-1))}{\partial\boldsymbol{u}^T(k-n_u-1)}\Delta\boldsymbol{u}(k-n_u) \\ +\boldsymbol{\varepsilon}_1\Delta\boldsymbol{u}(k)+\cdots+\boldsymbol{\varepsilon}_L\Delta\boldsymbol{u}(k-n_u) \tag{59}$$

and we let

$$\boldsymbol{\phi}_L^T(k)=[\frac{\partial\boldsymbol{f}(\boldsymbol{\varphi}(k-1))}{\partial\boldsymbol{u}^T(k-1)}+\boldsymbol{\varepsilon}_1(k),\cdots,\frac{\partial\boldsymbol{f}(\boldsymbol{\varphi}(k-1))}{\partial\boldsymbol{u}^T(k-n_u-1)}+\boldsymbol{\varepsilon}_L(k)] \tag{60}$$

to describe (59) as follows:

$$\Delta\boldsymbol{y}(k+1)=\boldsymbol{\phi}_L^T(k)\Delta\boldsymbol{U}(k) \tag{61}$$

, with $(\boldsymbol{\varepsilon}_1(k),\cdots,\boldsymbol{\varepsilon}_L(k))\to(\boldsymbol{0},\cdots,\boldsymbol{0})$ in nonlinear systems, when $(\Delta\boldsymbol{u}(k),\cdots,\Delta\boldsymbol{u}(k-n_u))\to(\boldsymbol{0},\cdots,\boldsymbol{0})$. As to linear systems, we will always have $\boldsymbol{\phi}_L^T(k)=[\frac{\partial\boldsymbol{f}(\boldsymbol{\varphi}(k-1))}{\partial\boldsymbol{u}^T(k-1)},\cdots,\frac{\partial\boldsymbol{f}(\boldsymbol{\varphi}(k-1))}{\partial\boldsymbol{u}^T(k-n_u-1)}]$, no matter what $(\Delta\boldsymbol{u}(k),\cdots,\Delta\boldsymbol{u}(k-n_u))$ is.

The proofs of Case 2: $1\le L\le n_u$ and Case 3: $L>n_u+1$ are similar to that of Case 1 and [21]. We omit the proofs.

We complete the proof of *Theorem* 1.

## VII. Appendix II

We define $\bar{\boldsymbol{\Psi}}(k)$ and $\boldsymbol{\Psi}(k)$ in the form of [22] as follows:

$$\bar{\boldsymbol{\Psi}}(k)=\begin{bmatrix} \boldsymbol{\phi}_L^T(k)\boldsymbol{A} \\ \sum_{i=0}^{1}\boldsymbol{\phi}_L^T(k+i)\boldsymbol{A}^{i+1} \\ \vdots \\ \sum_{i=0}^{N_u-1}\boldsymbol{\phi}_L^T(k+i)\boldsymbol{A}^{i+1} \\ \vdots \\ \sum_{i=0}^{N-1}\boldsymbol{\phi}_L^T(k+i)\boldsymbol{A}^{i+1} \end{bmatrix}$$

$$\boldsymbol{\Psi}(k) \\ =\begin{bmatrix} \boldsymbol{\phi}_L^T(k)\boldsymbol{B} & \boldsymbol{0} & \cdots & \boldsymbol{0} \\ \sum_{i=0}^{1}\boldsymbol{\phi}_L^T(k+i)\boldsymbol{A}^i\boldsymbol{B} & \boldsymbol{\phi}_L^T(k+1)\boldsymbol{B} & \vdots & \vdots \\ \vdots & \vdots & \ddots & \vdots \\ \sum_{i=0}^{N_u-1}\boldsymbol{\phi}_L^T(k+i)\boldsymbol{A}^i\boldsymbol{B} & \sum_{i=1}^{N_u-1}\boldsymbol{\phi}_L^T(k+i)\boldsymbol{A}^{i-1}\boldsymbol{B} & \ddots & \boldsymbol{0} \\ \vdots & \vdots & \ddots & \vdots \\ \sum_{i=0}^{N-1}\boldsymbol{\phi}_L^T(k+i)\boldsymbol{A}^i\boldsymbol{B} & \sum_{i=1}^{N-1}\boldsymbol{\phi}_L^T(k+i)\boldsymbol{A}^{i-1}\boldsymbol{B} & \cdots & \boldsymbol{\phi}_L^T(k+N-1)\boldsymbol{B} \end{bmatrix}$$

We define $\bar{\boldsymbol{\Psi}}(k,i)$, $\boldsymbol{\Psi}(k,i)$ and $\Delta\boldsymbol{U}(k,i-1)$ as follows: $\boldsymbol{\Psi}(k,i)=$

$$\begin{bmatrix} \boldsymbol{\phi}_L^T(k,i)\boldsymbol{B} & \boldsymbol{0} & \cdots & \boldsymbol{0} \\ \sum_{j=0}^{1}\boldsymbol{\phi}_L^T(k+j,i)\boldsymbol{A}^j\boldsymbol{B} & \boldsymbol{\phi}_L^T(k+1,i)\boldsymbol{B} & \vdots & \vdots \\ \vdots & \vdots & \ddots & \vdots \\ \sum_{j=0}^{N_u-1}\boldsymbol{\phi}_L^T(k+j,i)\boldsymbol{A}^j\boldsymbol{B} & \sum_{j=1}^{N_u-1}\boldsymbol{\phi}_L^T(k+j,i)\boldsymbol{A}^{j-1}\boldsymbol{B} & \ddots & \boldsymbol{0} \\ \vdots & \vdots & \ddots & \vdots \\ \sum_{j=0}^{N-1}\boldsymbol{\phi}_L^T(k+j,i)\boldsymbol{A}^j\boldsymbol{B} & \sum_{j=1}^{N-1}\boldsymbol{\phi}_L^T(k+j,i)\boldsymbol{A}^{j-1}\boldsymbol{B} & \cdots & \boldsymbol{\phi}_L^T(k+N-1,i)\boldsymbol{B} \end{bmatrix}$$

, $\Delta\boldsymbol{U}(k-1,i)=[\ \Delta\boldsymbol{u}^T(k-1,i),\cdots,\Delta\boldsymbol{u}^T(k-L,i)]^T$ ,

$$.\ \bar{\boldsymbol{\Psi}}(k,i)=\begin{bmatrix}\boldsymbol{\phi}_L^T(k,i)\boldsymbol{A}\\ \sum_{j=0}^{1}\boldsymbol{\phi}_L^T(k+j,i)\boldsymbol{A}^{j+1}\\ \vdots\\ \sum_{j=0}^{N_u-1}\boldsymbol{\phi}_L^T(k+j,i)\boldsymbol{A}^{j+1}\\ \vdots\\ \sum_{j=0}^{N-1}\boldsymbol{\phi}_L^T(k+j,i)\boldsymbol{A}^{j+1}\end{bmatrix}.$$